\documentclass{aa}
\usepackage{psfig}

\begin{document}

   \thesaurus{03         
              (03.09.2;  
               03.20.2;  
               05.01.1;  
               13.18.2)} 
   \title{A phase-reference study of the quasar pair 1038+528A,B }


   \author{Mar\'{\i}a J. Rioja 
          \inst{1}
          \inst{2}
          \and
          Richard W. Porcas\inst{3}
          }

   \offprints{Mar\'{\i}a J. Rioja}

   \institute{Observatorio Astron\'omico Nacional (OAN), Apdo. 1143, 
              E-28800 Alcal\'a de Henares, Spain \\
              email: rioja@oan.es
         \and
              Joint Institute for VLBI in Europe (JIVE), 
              NL-7990 AA Dwingeloo, The Netherlands \\
         \and
             Max-Planck Institut f\"ur Radioastronomie, Auf dem H\"ugel 69,
              D-53121 Bonn, Germany \\
             email: porcas@mpifr-bonn.mpg.de 
             }

   \date{Received 5 October, 1999; accepted XX }

   \titlerunning{Relative Astrometry of 1038+528AB}
   \authorrunning{Rioja \& Porcas}

   \maketitle
    
   \begin{abstract}

We present results from $\lambda$ 3.6~cm observations of the quasar pair 
\object{1038+528} A and B, made in 1995 using the VLBA together with the 
Effelsberg 100m telescope. We describe the use of a phase-referencing 
technique to measure the astrometric separation between the quasars.
We also introduce a new data analysis method - "hybrid double mapping" - 
which preserves the relative astrometric information in a single VLBI 
hybrid map for close source pairs. We combine our measurements with those 
from three previous epochs, the earliest in 1981. 
Our new observations confirm the evolution within the structure of quasar B, 
previously proposed to explain the measured change in the relative separation 
of the pair. Our upper bound for any systematic proper motion between the 
mass centres of quasars A and B is 10~$\mu$as~yr$^{-1}$. This is set by the 
limited precision in defining the reference points in the quasars at 
different epochs and by possible instabilities of the source "core" locations.
A separate analysis enables us to put more stringent upper limits to any 
core motions along the two source axes.

   \end{abstract}

%

\section{Introduction}

The quasar pair 1038+528~A,B (Owen et al. \cite{owen78}) 
consists of two
flat-spectrum radio sources, with redshifts 0.678 and 2.296 (Owen et
al. \cite{owen80}), separated on the sky by only 33\arcsec. 
This system provides a unique opportunity to carry out high precision, relative
astrometric studies using the full
precision of VLBI relative phase measurements, 
since most sources of phase errors are common for the 2 sources
(Marcaide \& Shapiro \cite{marca83}).

      VLBI studies of the mas-scale structure of flat-spectrum quasars
show that they typically have ``core-jet'' morphologies, consisting of a highly
compact feature (the ``core'') located at the base of an extended linear feature 
or line of lower brightness components (the ``jet'').
Both 1038+528 A and B exhibit such structures.
In standard models of extragalactic radio sources, these radio-emitting 
features
arise from a collimated beam of plasma which is ejected with a highly 
relativistic bulk velocity from a region close to a central massive object 
such as a black hole (see eg. Blandford \& K\"onigl \cite{blandford86}).
Whilst jet features may correspond to shocks in the moving plasma, and 
can give rise to the observed ``superluminal'' component motions in some 
sources (Porcas \cite{porcas87}),
the ``core'' emission is thought to arise from a more-or-less permanent location close to
the origin of the beam, where the ambient conditions correspond to a transition
from optically thick to optically thin emission at the observed frequency.
Although the ``core'' position may thus be frequency-dependent, for a fixed
observing frequency
the core should provide a stable marker, anchored to the central mass of
the quasar, whose location can be used to define a precise position for
the object as a whole.
Although short time-scale variations in physical conditions may cause
small changes in the ``core'' location, over long time-scales it may be
used to track any systematic proper motion of the quasar.

The results from a near decade-long VLBI monitoring program on 1038+52A,B at
$\lambda$ 3.6 and 13~cm (from 1981.2 to 1990.5) 
are reported by Rioja et al. (\cite{rioja97}), whose main conclusions 
can be summarized as follows:

\begin{enumerate}

\item There is no evidence of any relative proper motion between the 
      quasars A and B. The uncertainties in the astrometric parameters 
      result in an upper bound to any systematic relative 
      motion between the cores of 10~$\mu$as~yr$^{-1}$, consistent
      with zero.

\item A compact feature within the jet of quasar B, chosen as the 
      reference point for the structure, expands away from the core at 
      a steady, slow rate of $\sim~18~\pm~5~\mu$as~yr$^{-1}$, corresponding 
      to v=$(0.8~\pm~0.2)~h^{-1}$~c for a Hubble constant,
      H$_0$ = 100~h~km~s$^{-1}$~Mpc$^{-1}$; q$_0=0.5$. These values are used 
      here throughout.

\item The accuracy of the relative separation measurement is limited by 
      noise and source structure, with estimated precisions of about 
      $50~\mu$as at $\lambda$ 3.6~cm at any epoch.

\item Confirmation of the consistently large offset (about 0.7 mas) between
      the positions of the peak of brightness (``core'') at $\lambda$ 3.6 
      and 13~cm in quasar A.

\end{enumerate}

New VLBI observations of this pair were made in November 1995 (1995.9) 
at $\lambda$ 2, 3.6 and 13~cm. In this paper we report on results from 
our analysis of the 3.6~cm observations and investigate the temporal 
evolution of the source structures and relative separation from all 
four epochs spanning $\sim 15$ years. 
Investigations of frequency-dependent source structure have also been made 
from a comparison of the astrometric measurements
of the separations between A and B at all 3 wavelengths observed in 1995;
these will be presented elsewhere (Rioja \& Porcas in preparation).

Our new observations are described in Sect. 2. In Sect. 3 we describe
the data reduction and mapping techniques used, and in Sect. 4 an analysis
of the measurements in the maps.
In Sect. 5 we compare the astrometric results from these observations
with those from previous epochs and analyse the changes in 
separation. Conclusions are presented in Sect. 6.


\section{Observations}

The pair of radio sources 1038+528 A and B was observed with the
NRAO Very Long Baseline Array (VLBA) on November 10, 1995,
for a total of 13 hours, alternating every 13 minutes between
observations in dual 3.6/13~cm mode and observations at 2~cm.  The
100m telescope at Effelsberg was also included in the array for the
3.6/13~cm scans. The primary beamwidths of all the antennas were
sufficiently large that both sources could be observed simultaneously
at all wavelengths. Each 10 minute observation of 1038+52A,B was
preceded by a 3 minute observation of the compact calibration source
\object{0917+624}, to monitor the behaviour of the array.

All stations used VLBA terminals to record an aggregate 
of 64 MHz bandwidth for each scan, using 1-bit sampling, subdivided into
8 channels (mode 128-8-1). For the dual 3.6/13~cm scans, four 8-MHz channels
were recorded for each band (2254.5--2286.5 MHz; 8404.5--8436.5 MHz), 
using 
RHC polarisation. At 2~cm, eight 8-MHz channels (15\,331.5--15\,395.5 MHz)
were recorded in LHC polarisation.

The correlation was made at the VLBA correlator in Socorro (New Mexico).
As for previous epochs, two separate ``passes'' were needed, using different 
field centres for the two sources, to recover data
for both the A and B quasars from the single observation. 
Output data sets were generated for the two sources, consisting of the
visibility functions averaged to 2 s, with samples every 1 MHz in frequency
across the bands. 

\section{Data reduction} 

We used the NRAO AIPS package for the data reduction. 
We applied standard fringe-fitting, amplitude and phase (self-) calibration 
techniques and produced hybrid maps of each quasar. 
The astrometric analysis was done using two different mapping methods: 
a ``standard'' phase-referencing approach, transferring phase solutions
from one quasar to the other
(see e.g. Alef \cite{alef88}; Beasley \& Conway \cite{beasley95}) and a 
novel mapping method for astrometry of close pairs
of sources, hybrid double mapping (HDM) (Porcas \& Rioja \cite{porcas96}).  
Both routes preserve the signature of the relative separation of the source 
pair present in the calibrated phases. These analysis paths are described 
in Sects. 3.1 to 3.3 below.   

\subsection{Hybrid mapping in AIPS} 

We applied standard VLBI hybrid mapping
techniques in AIPS for the analysis of the observations of both
quasars A and B. We used the
information on system temperature, gain curves and telescope gains measured at
the individual array elements, to calibrate the raw correlation
coefficients. 
We used the AIPS task FRING to estimate residual antenna-based phases
and phase derivatives (delay and rate) at intervals of a few minutes.
It is important to realise that FRING is a global self-calibration
algorithm, and performs an initial phase self-calibration also. We ran
FRING on the A quasar data set, with a point-source input model.

Anticipating our phase-referencing scheme (Sect. 3.2) we applied the antenna
phase, delay and rate solutions from A to both the A and B data sets, and
averaged them in time to 60 s, and over the total observed bandwidth of 32 MHz.
After suitable editing of the data, we made hybrid maps of both quasars,
using a number of iterations of a cycle including the mapping task MX and
further phase self-calibration with CALIB.

\begin{figure*}
\centerline{
{\psfig{figure=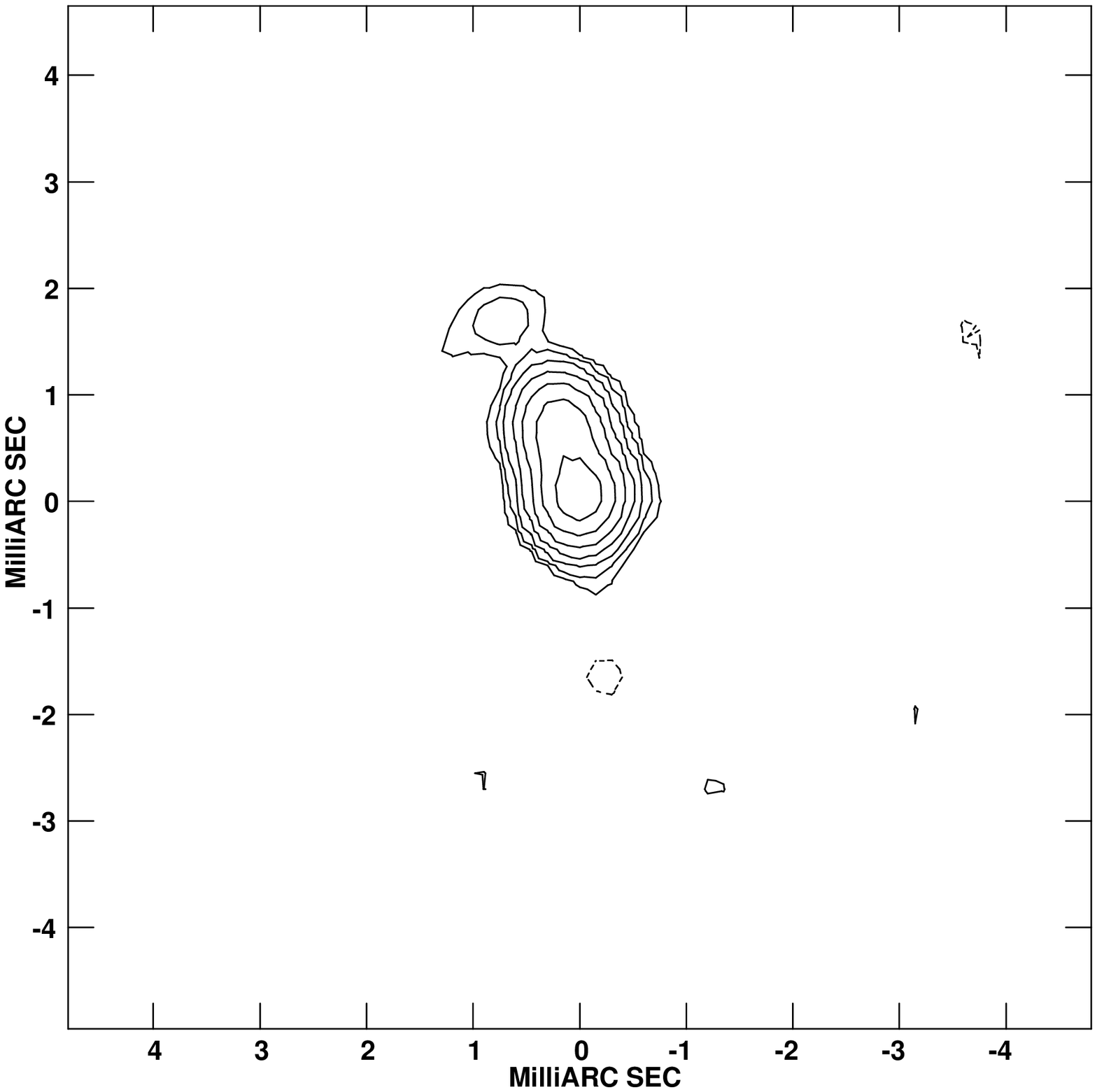,width=6.cm}}
\put (-150,150) {\bf {a)}}
{\psfig{figure=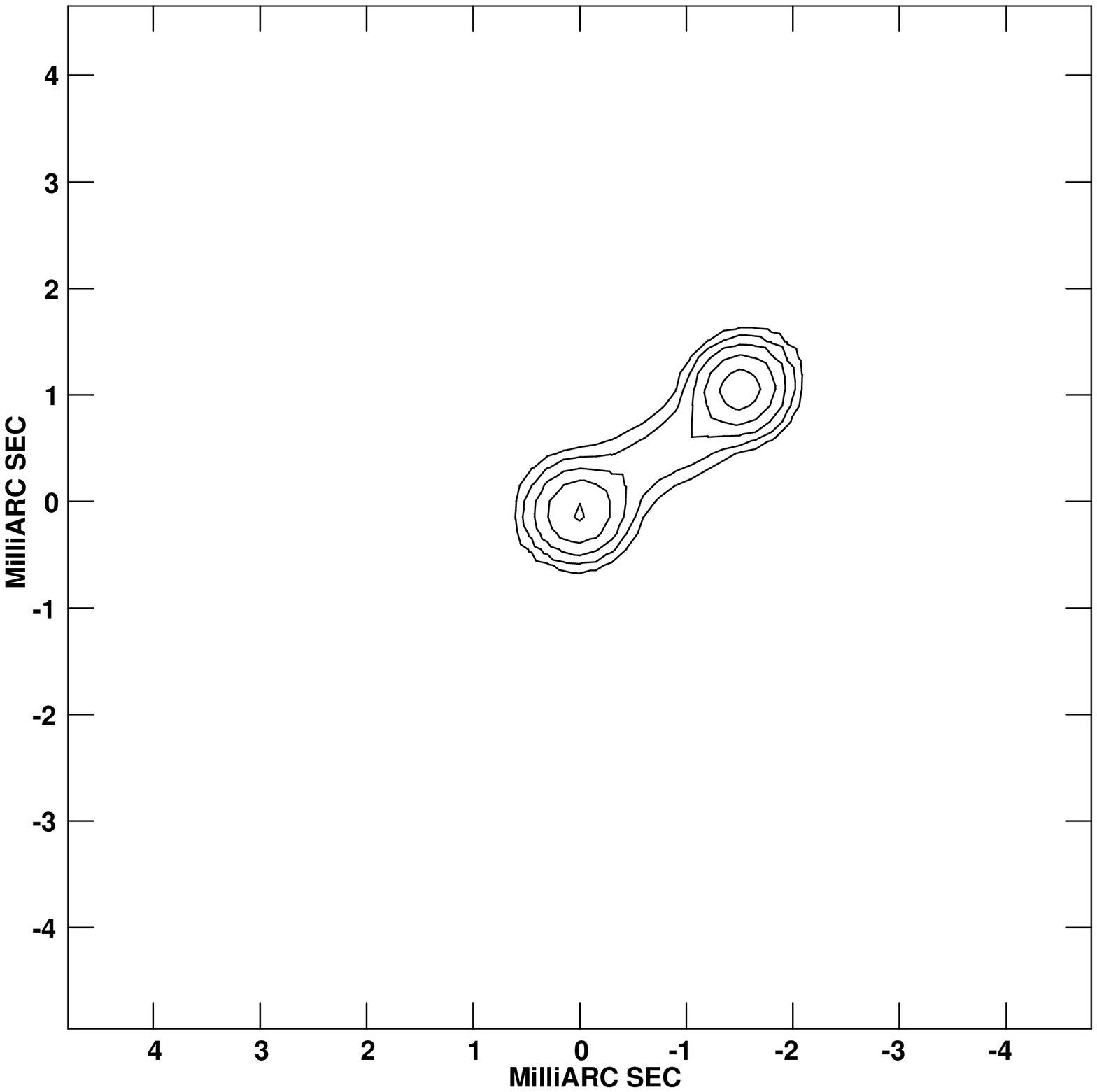,width=6.cm}}
\put (-150,150) {\bf {b)}}
}
\caption{VLBI hybrid maps of 1038+528 at 3.6~cm. Uniform weighting, 
CLEAN beam 0.5 x 0.5 mas, pixel size 0.15 mas, tick interval 1 mas.
{\bf a} Quasar A. Contours at 3,6,12,24... mJy/beam.
{\bf b} Quasar B. Contours at 1.5,3,6,12... mJy/beam.
\label{fig1}} 
\end{figure*}

Fig.~\ref{fig1}a and b show the hybrid maps for both sources at 3.6~cm 
in 1995.9. 
The maps are made using uniform weighting of the visibilities, a map
cell size of 0.15 mas and a circular CLEAN restoring beam of 0.5 mas
(these same mapping parameters are used throughout this work).
The ``dirty'' beam has a central peak of 0.57 x 0.47 mas in PA -29\degr
(PA = position angle, defined starting at North, increasing through East).
The root-mean-square (rms) levels in the A and B maps, in regions away 
from the source structures (estimated using AIPS task IMSTAT) are 1.0 and 
0.12 mJy/beam respectively, an indication that dynamic range considerations 
dominate over thermal noise in determining the map noise levels.  


\subsection{Phase referencing in AIPS}

In order to make an astrometric estimate of the separation between
quasars A and B at this 4th epoch, we first used a "conventional" 
phase-reference
technique to make maps of the quasars which preserve the relative phase
information.
In practice this consists of using the antenna-based residual terms 
derived from the analysis of the data of one ``reference'' source (A), 
to calibrate the data from simultaneous observations of the other "target" 
source (B).
The reference quasar source structure must first 
be estimated from a hybrid map, and then fed back into
the phase self-calibration process to produce estimates of the antenna-based  
residuals, free from contamination by source structure. 

Phase referencing techniques work under the assumption that the
angular separation between the reference and target sources is smaller
than the isoplanatic patch size (i.e. the effects of unmodelled perturbations,
introduced by the propagation medium,
on the observed phases of both sources are not very different) and that 
any instrumental terms are common. Geometric errors in the correlator model
must also be negligible.

Assuming that the antenna residuals have been ``cleanly'' estimated using the
reference source data, the calibrated phases of the target 
source should be free from the errors mentioned above, but still retain the
desired signature of the source structure and relative position
contributions. The Fourier Transformation of the calibrated visibility
function of the target source produces a ``phase referenced'' map. The
offset of the brightness distribution from the centre of this map reflects
any error in the assumed relative separation in the correlator model.
If the reference source has a true ``point'' structure and is at the
centre of its hybrid map, this offset will be equal to the error;
more generally, one should also measure the offset of a
reference point in the reference source map, and estimate the
error in the source separation used in the correlator model from the difference
between the target and reference source offsets.

In general, the success of the phase-referencing technique is critically
dependent on the angular separation of the target and reference sources.
Simultaneous observation of the sources, as was possible here, significantly 
simplifies the procedure, eliminates the need for temporal interpolation,
and reduces the propagation of errors introduced in the
analysis.  While random errors increase the noise level in the phase
referenced map, systematic errors may bias the estimated angular
separation.

\begin{figure*}
\centerline{
{\psfig{figure=fig2a.ps,width=6.cm}}
\put (-150,150) {\bf {a)}}
\put (-84.5,88.5) {\bf {+}}
{\psfig{figure=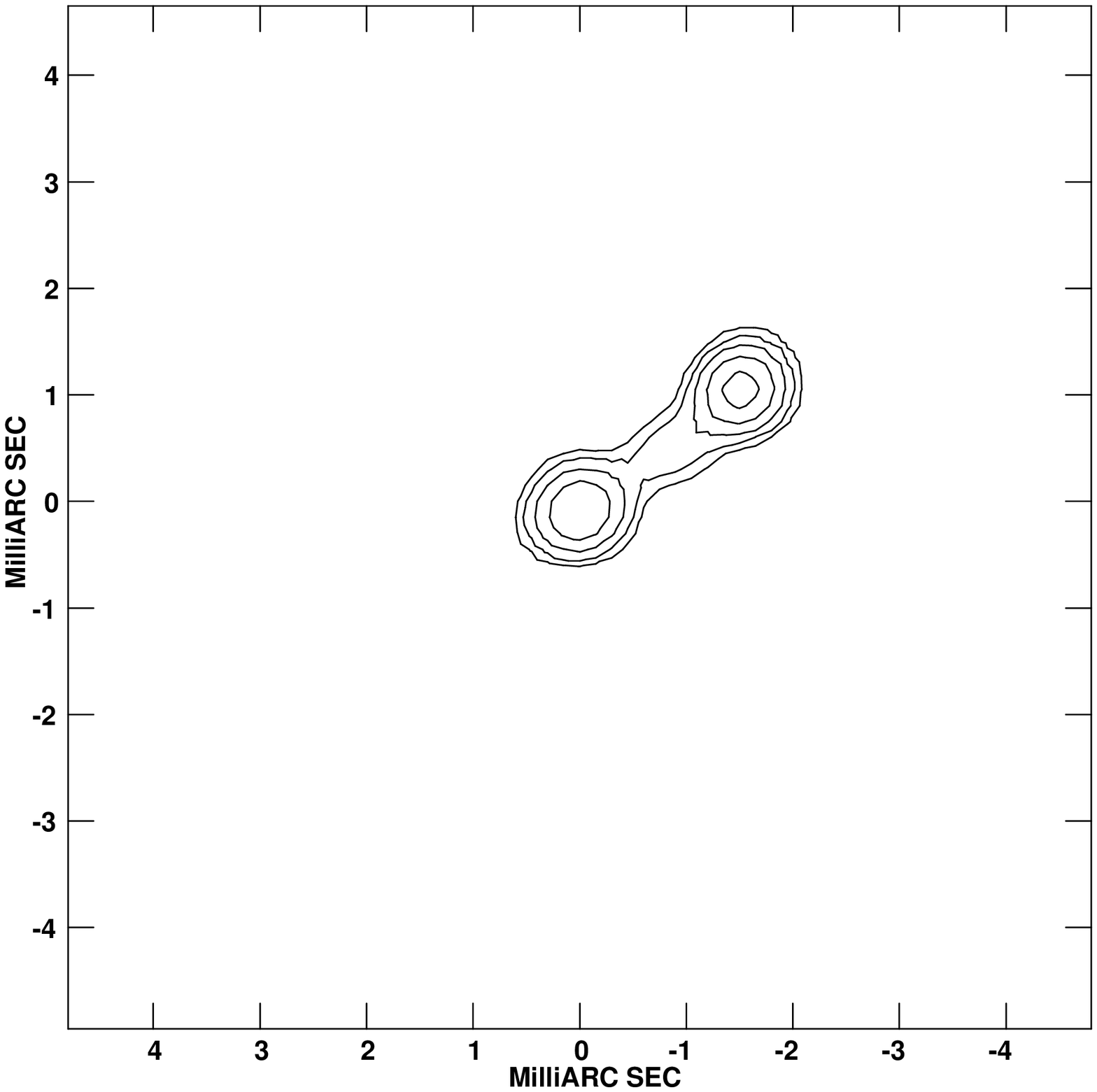,width=6.cm}}
\put (-150,150) {\bf {b)}}
\put (-85,88) {\bf {+}}
}
\caption{VLBI phase-reference astrometry maps. Map parameters as in Fig. 1.
Astrometry reference points are indicated with a cross.
{\bf a} Quasar A. Contours at 3.5,7,14,28... mJy~beam$^{-1}$.
{\bf b} Quasar B. Contours at 1.5,3,6,12... mJy~beam$^{-1}$.
\label{fig2}} 
\end{figure*}

For our implementation of phase-referencing using AIPS, we chose
to re-FRING the (calibrated) A data set, using our hybrid map of quasar 
A as an input model, and applied the adjustments to the antenna phase, 
delay and rate solutions to both the A and B data sets before re-averaging. 
We then made maps of both A and B using MX, performing no further phase 
self-calibration. These are our ``phase-reference astrometry'' (PRA) 
maps (shown in Fig.~\ref{fig2}a and b) on which we performed
astrometric measurements (see Sect. 4.2). Although the rms noise levels 
in the PRA maps are slightly higher than in the corresponding hybrid maps 
(2.0~mJy~beam$^{-1}$ for A and 0.24~mJy~beam$^{-1}$ for B),
our procedure ensures that the A and B
visibility functions from which they are derived have been calibrated 
identically. 


\subsection{New mapping method for astrometry of close source pairs} 

While the conventional phase-referencing approach worked well for our
November 1995 observations of 1038+52A and B, the method relies on
making a good estimate of the antenna residuals from just one of the 
sources - the reference. We have devised an alternative method which
extends the standard VLBI self-calibration procedure to work on both
sources together, for cases where they have been observed simultaneously,
and when either could be used as the reference (see Appendix A).

The basis of the new method is to recognise that, since the visibility 
functions for both sources are corrupted by the same (antenna-based) 
phase and phase derivative errors, the sum of the two visibilities also 
suffers the same errors. We form the point-by-point sum of the two data 
sets, creating a new one which represents the visibility function of a 
``compound source'' consisting of a superposition of the two structures, 
corrupted by the common antenna phase errors.
If the source separation is close enough, the (summed) data as
a function of the (averaged) uv-coordinates can be
Fourier Transformed to form a map of the
compound source structure, and (iterative) self-calibration in FRING or CALIB
yields the antenna-based residuals. The advantage of this approach is that
the antenna-based residuals are determined using both source structures
simultaneously, and may thus reduce the chance that reference source structural
phase terms contaminate the residuals. We term this process ``Hybrid Double
Mapping'' (HDM); a detailed description is given in Porcas \& Rioja 
(\cite{porcas96}).

It is convenient to shift the source position in one of the data
sets (by introducing artificial phase corrections)
prior to the combination into a compound-source data set, to avoid  
superposition of the images in the map. The phase self-calibration 
steps which are then applied to the combined data set are 
identical to the case of a single source. In HDM the information
on the angular separation between the sources is preserved in the 
process of self-calibration of the combined visibilities, and can be measured
directly from the compound-image map;
the relative positions between the individual source images in the 
compound map, taken together with any artificial position shift introduced,
give the error in the assumed angular separation in the correlator model. 
In this approach one must be careful to use the same number of visibility
measurements in each time interval from the two data sets, in order to avoid
the predominance of data from a particular source.  

\begin{figure}[h]
\centerline{
{\psfig{figure=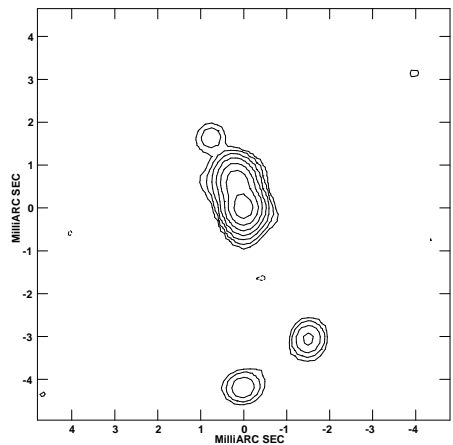,width=6.cm}}
}
\caption{HDM map of 1038+52 (A+B). Map parameters as in Fig. 1.  
Quasar B has been offset by -4.0 mas in
declination. Contours at 3,6,12,24... mJy~beam$^{-1}$. 
\label{fig3}} 
\end{figure}

Fig.~\ref{fig3} shows the HDM map of quasars A and B in 1995.9 at 
$\lambda$ 3.6~cm; 
the B source is artificially offset by -4 mas in declination.
The rms noise in the map is 0.82~mJy~beam$^{-1}$ - higher than that in 
the hybrid map of B but lower than in that of A.


\section{Analysis of the maps} 

\subsection{Source structures} 

The 1995.9 hybrid maps of quasars A and B at $\lambda$ 3.6~cm 
(Fig.~\ref{fig1}) show the core-jet structures typical of quasars at 
mas scales. They may be compared with maps from previous epochs given 
in Rioja et al. (\cite{rioja97}). 
The structure of A in the new map shows no major changes with respect 
to previous epochs. There is a prominent peak at the SW end of the
structure (the ``core'') and a jet extending in PA 15--25\degr
~containing at least two ``knot'' components (k1 and k2).

The new map of B at 3.6~cm is qualitatively similar to
those from previous epochs. It shows 2 point-like components separated
by just under 2 mas in PA 127\degr.
Spectral arguments support the identification of
the NW component as a ``core'' (Marcaide \& Shapiro \cite{marca84}); 
the SE component, corresponding to a knot in the jet, has been used as 
a reference component in previous astrometric studies.
The separation between these 2 components in 1995.9 has increased, 
continuing the expansion along the axis of the source, as discovered 
from previous epochs of observations at this wavelength (Rioja et al. 
\cite{rioja97}).
There is no trace of the third, extreme SE component, seen in maps
of this source at 13~cm. This feature is evidently of lower surface
brightness at 3.6~cm and is resolved out at the resolution of these 
observations.

We used AIPS task IMSTAT to estimate total flux densities for quasars A and
B (within windows surrounding the sources in the hybrid maps). The values 
are given in Table~\ref{tab1}.
Table~\ref{tab1} also lists the fluxes and relative positions of the 
most prominent features in the maps of A and B, obtained using task JMFIT to
find parameters of elliptical Gaussian functions which best
fit the various source sub-components. The formal errors from the fits,
however, do not give realisitic values for the parameter uncertainties.
The distribution of flux between the core and k1 in quasar A, and their
relative separation, are quite uncertain, for example.  

\begin{table*}
\caption {Parameters derived from the Hybrid Maps of 1038+528 A and B.} 

\vspace*{0.25cm}

\begin{center}
\begin{tabular} {|c|c|c|c|c|c|c|c|c|c|} \hline
   &  S-tot & &  S-pk  & S-int & maj.& min. & PA & sep. & PA \\
   &  (mJy) & &(mJy/b) & (mJy) & (mas) & (mas) & (deg) & (mas) & (deg) \\
\hline
1038+528 A & 603.0 & core & 301.6 & 355.3 & 0.28 & 0.12 & 3 &   - & - \\
           &       & k1 & 171.8 & 235.5 & 0.42 & 0.16 &   13 & 0.639 & 15.1 \\
           &       & k2 &   9.3 & 14.0 &  -   &   -  &    - & 1.796 & 24.8 \\
&&&&&&&&& \\
1038+528 B & 86.5 & core & 33.5 & 40.9 & 0.32 & 0.11 & 132 &  -  & - \\
         &      &  ref & 24.6 & 34.4 & 0.39 & 0.24 & 123 & 1.869 & 127.1 \\ 
\hline \hline
\end{tabular}
\end {center}
\label{tab1}
\end{table*}

\subsection{Estimating positions of reference features} 

The astrometric measurement of a separation between two non-point sources
must always refer to the measured positions of reference points within 
maps (or other representations) of the source structures. 
The selection of suitable reference points is crucial in monitoring
programs, where the results from the analysis of a multi-epoch  
series of observations are compared. 
Ideally, a reference point should correspond to the peak of
a strong, unresolved component, which is well separated from
other radio emission within the source structure.

For the 1995.9 epoch observations of 1038+528 A,B we selected the same 
reference features as those used for the analyses of previous observing 
epochs. These are the ''core'' component for quasar A, and the prominent 
SE component for quasar B.
These features are labelled with a cross in Fig.~\ref{fig2}a and b.
The core of A is indeed strong and compact, but has the disadvantage
that it merges with knot k1.
Although the SE component of B is no longer the strongest feature at 3.6~cm,
it has always been strong at both 3.6 and 13~cm wavelengths,
is reasonably compact and is easily distinguishable
in maps made at longer wavelengths, thus facilitating spectral studies.
Our astrometric analysis refers to the measured positions of the peaks
of these components in A and B. We used the AIPS task MAXFIT to measure 
the position of these peaks in the PRA and HDM maps.
MAXFIT defines the location of a peak in a given map region by fitting 
a quadratic function to the peak pixel value and those of the adjacent pixels.
A comparison of this method of defining the peak position with that used
for earlier epochs is described in the next section. 

\subsection{Position error analysis} 

An analysis of errors presented in Rioja et al. (\cite{rioja97}) shows 
that the dominant uncertainty in the astrometric measurements of the 
separation between this close pair of quasars comes from the limited 
reproducibility of the reference point positions in the VLBI maps, 
from epoch to epoch. The magnitude of this effect
is hard to quantify, however, since it depends on the nature of the
source brightness distribution surrounding the reference point, and
the method used to define the position of the peak, in addition
to the resolution of the array and the signal-to-noise
ratio of the peak in the map.

A rough estimate of the error due solely to finite signal-to-noise in
the maps is given by dividing the beam size by the ratio of the
component peak to the rms noise level in the maps (see e.g. 
Thompson et al. \cite{thompson86}). 
This yields values of 3.3, 3.5 and 1.4~$\mu$as for the A and B PRA 
maps and the HDM map. These may be taken to represent lower limits to 
the reference point position errors; realistic errors will be larger,
and will depend on the nature of the reference features 
and the manner in which the position is estimated.

It is important to choose a definition of the reference point position 
such that it can be reproduced reliably from epoch to epoch, and is as 
independent as possible from the parameters used in making the map 
(e.g. cell size and beam width).
The AIPS task JMFIT can be used to fit an elliptical Gaussian to a component
in a CLEAN map, for example. However, the position of the peak of the 
Gaussian depends on how asymmetric the component brightness distribution 
is, and the area of the map to which the fit is restricted. MAXFIT fits 
just to the local maximum around the peak map value, and is thus less 
sensitive to the rest of the distribution.

We have attempted to quantify some limits to reproducibility arising from 
the use of MAXFIT for defining the peak position in CLEAN maps.
We investigated the effect of changing the true position of a point-like
source with respect to the pixel sampling (here 3.3 pixels per beam) 
by offsetting the source position in 10 increments of 1/10 of a pixel 
in the visibility domain, mapping and CLEANing the new data sets, and 
estimating the new positions in the CLEAN maps using MAXFIT. The
maximum discrepancy found between the values of the artificial offset
and the shift derived by MAXFIT was 1/20 of a pixel.
This corresponds to 8~$\mu$as in our 3.6~cm maps.

For the analysis of previous observing epochs, the reference points
were defined to be the centroid of the most prominent delta functions 
from which the CLEAN source map was derived (Rioja et al. \cite{rioja97}).
We examined possible systematic differences resulting from these
different definitions of reference points.
One might expect the largest discrepancies to arise when the underlying
source structure near the reference point is asymmetric, as in quasar A.
We investigated such differences by determining ``centroid'' positions for
both A and B reference components, using various criteria for excluding
clean components from the calculation; this included the ``25 percent
of the value at the peak'' threshold used for earlier epochs.
For A the difference between this centroid position and the MAXFIT
value was 0.12 pixel (18~$\mu$as). For B the difference was less
than 0.1 pixel.
These are probably the largest potential sources of error arising from using
different methodologies at different epochs.

Our use of two different mapping procedures - phase-reference mapping
and HDM - also gives some insight into the size of position errors resulting
from standard CLEAN + phase self-cal mapping algorithms.
The differences between the separation estimates from the PRA maps and the
HDM map are 27 and 28~$\mu$as in RA and Dec respectively.
This would suggest that differences in the positions of peaks in 
maps reconstructed in different ways may vary at the 14~$\mu$as level.

After considering the various possible effects which can limit the
accuracy of postion estimates, we adopt a ``conservative'' value for 
the error in estimating the peak position in our $\lambda$ 3.6~cm maps,
embracing all the effects detailed above, of $18~\mu$as
(this corresponds to a thirtieth of the CLEAN beam).
The associated estimated error for a separation measurement between
the two sources is $25~\mu$as.

\subsection{Astrometry results} 

Table~\ref{tab2} lists the results of our astrometric measurements of 
reference point positions in the maps.
They are presented as changes in measured separation between the 
reference points in A and B in 1995.9, with respect to their separation 
in 1981.3. 
The values given from the phase-reference technique correspond to the 
difference between the A and B reference feature position offsets in 
their respective PRA map. The values derived from HDM have been corrected 
for the artificial offset introduced before adding the A and B source 
visibilities.


\begin{table}
\caption{Change in the separation between quasars A and B in 1995.9
with respect to 1981.2, estimated using standard phase referencing 
(PRA) and hybrid double mapping (HDM) techniques.}
\begin{center}

\vspace*{0.25cm}

\begin{tabular} {|c| c| c|} \hline 
$\Delta(\Delta \alpha)$ $\cos \delta_A$ & $\Delta(\Delta \delta)$ & Method \\ 
$[\mu$as$]$ & $[\mu$as$]$ & \\ \hline \hline
 -148 & 249 & PRA  \\
 -175 & 277 & HDM \\ \hline \hline
\end{tabular}
\end{center}
\label{tab2}
\end{table}

All these values have been corrected for a small error in the AIPS calculation
of the u,v coordinates in the frequency-averaged data set.
(Distances measured within the maps must be adjusted
by a small correction factor of
$1 - \Delta \nu \ast (2~\nu)^{-1} = 0.998$.) 
 \\

Table~\ref{tab3} lists the coordinates of the reference source (A) adopted in
the analysis and the measured coordinate separation between quasars A
and B in 1995.9.


\begin{table*}
\caption{{\it Fixed} source coordinates used for quasar A
in the astrometric analysis (these coordinates correspond to GSFC global
solution GLB831 (Chopo Ma, priv. comm.)), and
separation between quasars A and B measured in 1995.9.}

\begin{center}

\vspace*{0.25cm}

\begin{tabular} { |l| r| r|} \hline
{\bf Coordinates} & {\bf RA} (J2000) \hspace*{0.5cm} & {\bf DEC} (J2000) \hspace*{0.5cm}
 \\ \hline \hline
{\it Reference Source {\bf (A)}}  & 10$^h$ 41$^m$ 46\fs781613 
\hspace*{0.05cm} & 52$^0$ 33\arcmin 28\farcs23373 \hspace*{0.05cm} \\ \hline 
{\it Relative Sep. {\bf (B-A)}} & $2\fs1160588  $ 
& $27\farcs376325$ \\
{\it Estimated error {\bf (B-A)}} & $\pm 0\fs0000027$ & $ \pm 0\farcs000025$ \\ \hline
\end{tabular}
\end{center}
\label{tab3}
\end{table*}

\section{Comparison of astrometry at all 4 epochs} 

In this section we make a comparison of the astrometric
measurements from the series of 4 epochs of observations.
Any increase of the temporal baseline in the program of monitoring the 
separation between A and B should result in a more precise 
identification of any systematic trends,
with an improved elimination of random contributions.
In Sect. 5.1 we justify comparing
the astrometric values measured at the various epochs, even
though non-identical observing, post-processing and analysis
procedures were involved.
In Sect. 5.2 we present the astrometric results from the 4 epochs.
In Sects. 5.3, 5.4, 5.5 and 5.6 we present various analyses of these
results, and attempt to quantify, or put upper-limits to,
proper motions within and between the A and B quasars.

\subsection{Comparison between the techniques used at different epochs}

Before attempting a comparison of the astrometric results from 
the 4 observing epochs, we need to
show that any bias in the astrometric estimates introduced by the
use of different procedures is small compared with other errors in
the measurements for the individual epochs.
The consistency between the results from previous epochs of observations 
has been exhaustively tested (Marcaide et al. \cite{marca94}; 
Rioja et al. \cite{rioja97}).
We outline here the largest changes involved in the fourth epoch, 1995.9,
with respect to previous ones:

\begin{enumerate}

\item The observing array and frequency set-up used in the fourth epoch
was different from previous epochs of observations 
(frequency range 8404.5 to 8436.5 MHz instead of 8402.99 to 8430.99 MHz
at first 3 epochs). This results in a different coverage of the {\it UV} 
plane, leading to changes in the reconstruction of the source images. 
Investigations of such effects by Marcaide et al. (\cite{marca94}) show that
the effect on the astrometric anaylsis is only a few $\mu$as. 
It is important to note that the observations at all 4 epochs have
comparable resolutions and sample the same range of structural scales
in the sources.

\item The processing of the fourth epoch was done using the VLBA correlator,
which uses a theoretical model derived from CALC 8.2; we used AIPS to
analyse the data with visibility phases residual to that model.
For previous epochs the correlation was done at the MPIfR (Bonn) MK3
correlator, and an analysis of the data using total phases was made
with VLBI3 (Robertson \cite{robertson75}).
The differences between CALC 8.2 and the one implemented in VLBI3 
propagate into changes of only 1-2~$\mu$as in the astrometric analysis
of the 1038+528 A-B separation (Rioja \cite{rioja93}, Rioja et al. 
\cite{rioja97}). 
This is because any such differences are ``diluted'' by the source 
separation expressed in radians - 
$10^{-4}$ in the case of this very close source pair.

\item The values used in the analysis of previous epochs for
Earth Orientation Parameters (EOP), stations and reference source coordinates
were consistently derived from a single global solution provided by Goddard 
Space Flight Center (GSFC). 
For the correlation of the fourth epoch, the values used for 
EOP were derived from IERS solutions, and the station coordinates from 
USNO catalogs.
We have made a comparison of the values derived for all the parameters at
the 4 epochs from a single global solution from IERS (namely IERS eopc04),
with the actual values
used in the individual epoch analysis. The difference between the corresponding
EOP values is always less than 4 mas. Such discrepancies propagate into 
errors in the relative position estimates at each epoch of only a few 
$\mu$as.

\item Our astrometric analysis in AIPS using a phase-referencing approach and
HDM differs from the phase difference method used in VLBI3 analysis.
Comparisons show that these procedures are equivalent 
(Porcas \& Rioja \cite{porcas96}; Thompson et al. \cite{thompson86}).
Both involve the definition of reference points in source maps;
uncertainties in the reference point positions 
(as described in Sect. 4.3) arise in the same way.

\item Finally, a minor VLBA correlator error (Romney priv. comm.) caused
incorrect time labels to be attached to the visibility records, resulting in
incorrect ({\it u,v}) values. 
The effect on the relative visibility phases for our source pair is small
($\sim 0.004$\degr) and can be neglected.

\end{enumerate}

   The magnitudes of all of the effects reported in this section are
much smaller than our estimate in Sect. 4.3 of the uncertainity in reproducing
the reference point in the source, from epoch to epoch, and we are thus
justified in comparing the astrometric results from all 4 epochs.

\subsection{Astrometric separations at the 4 epochs} 

\begin{figure}[h]
\centerline{
{\psfig{figure=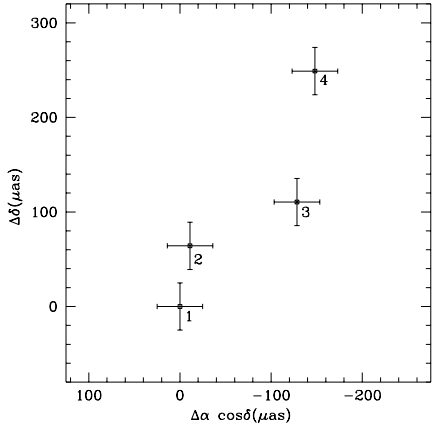,width=6.cm}}
}
\caption{Measured separations between the A and B reference points at epochs
2 (1983.4), 3 (1990.5) and 4 (1995.9), with respect to epoch 1 (1981.2).
Plotted error bars correspond to $25~\mu$as.
\label{fig4}} 
\end{figure}

The astrometric measurements of the separations between the reference 
points in A and B at $\lambda$ 3.6~cm from 4 epochs are presented in 
Fig.~\ref{fig4}.
It includes our new 1995.9 measurement and those from three earlier epochs, 
in 1981.3, 1983.4 and 1990.5, reported in Marcaide \& Shapiro 
(\cite{marca84}), Marcaide et al. (1994) and Rioja et al. (\cite{rioja97}), 
respectively. The origin of the plot represents the separation at epoch 1.


Changes with time in Fig.~\ref{fig4} represent the vector difference between
any motions of the reference points in quasars A and B.
The near-orthogonal nature of the source axes in 1038+52 A,B (along
which one might expect any motion to occur) simplifies the
interpretation of any trends seen. The new 1995.9 value follows the
same steady trend towards the NW shown by the three previous epochs.
Rioja et al. (\cite{rioja97}) interpreted this as an outward expansion of the
reference component in quasar B at a rate of $18~\pm~5~\mu$as~yr$^{-1}$, 
and quoted an upper bound on any proper motion of quasar A of 
$10~\mu$as~yr$^{-1}$.

\subsection{Vector decomposition} 

In this section we attempt to separate the individual contributions from 
the 2 quasars in the astrometric separation measurements presented in 
Fig.~\ref{fig4}. We make no assumption about the stability of either 
component, but
assume that any displacements of the A or B reference points from their 
positions at epoch 1 are along the corresponding source axis directions.
This is a plausible assumption if the reference point coincides 
with a non-stationary component moving along a ballistic trajectory, 
or with the location of the peak of brightness within an active core 
or near the base of jet, where changes during episodes of activity are 
likely to occur along the jet direction.
This approach is closely related to that used previously 
by Rioja et al. (\cite{rioja97}). For fixed assumed source axes for A and B,
it results in a unique decomposition of the changes in the A-B separation
into separate A and B displacements, from 1981 to 1995.

It is clear that the dominant contribution to the separation changes
seen in Fig.~\ref{fig4} comes from quasar B, 
in which the source axis is well defined by the 127\degr PA of
the separation between core and reference components.
For quasar A the source axis bends, from the inner ``core'' region 
(PA = 15\degr) to the outer jet components, and it is not so clear which 
direction should be chosen.

\begin{figure}[h]
\centerline{
{\psfig{figure=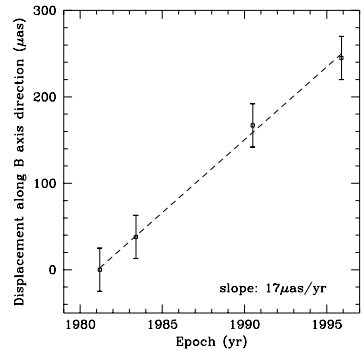,width=6.cm}}
}
\caption{Changes in position of reference component in B along PA 127\degr,
deduced from deconvolution of the A-B separation measurements. Assumed source
axis for A is 25\degr.
Plotted error bars correspond to $25~\mu$as.
\label{fig5}} 
\end{figure}

In our analysis we tried a range of values for fixing the A source axis 
(0 to 45\degr ~in steps of 5\degr).
For each, we calculated A and B reference-point displacements at epochs 2, 
3 and 4 with respect to epoch 1.
Then we performed a least-squares fit to the B displacements with time to
estimate a linear expansion rate for the B reference feature along 
PA 127\degr. In Fig.~\ref{fig5} we plot the deconvolved B reference point 
displacements from the analysis with the A source axis fixed at 
PA 25\degr (the value adopted by Rioja et al. \cite{rioja97}).
The fitted expansion rate is $16.9~\pm~0.6~\mu$as~yr$^{-1}$;
the error and associated rms values take account of the small number 
of points and 2 degrees of freedom. This rate agrees well with the value
of $18~\pm~5~\mu$as~yr$^{-1}$ deduced by Rioja et al. (\cite{rioja97}).
The rms residual from the fit (7~$\mu$as) is low, and vindicates
our use of measurements derived from differing techniques for
investigating the relative proper motion between A and B. 


\subsection{Structural evolution within 1038+528 B} 

Our deconvolution analysis of the changes in separation measured between 
all 4 epochs supports the finding, previously proposed, that the B reference
component moves along the source axis, away from the B core.
In this section we make an independent determination of the separation 
rate between the core and reference component in B from measurements 
within the maps at the 4 epochs.

\begin{figure}[h]
\centerline{
{\psfig{figure=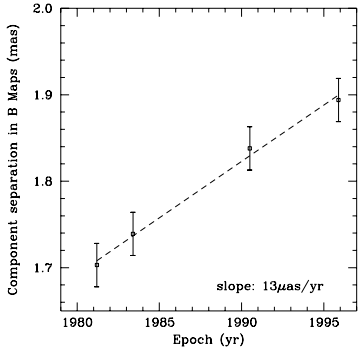,width=6.cm}}
}
\caption{Changes in the position of the reference component in B 
along PA 127\degr, from measurements of its separation from the core
in hybrid maps of B.
Plotted error bars correspond to $25~\mu$as.
\label{fig6}} 
\end{figure}

Fig.~\ref{fig6} shows the separation between the core and reference 
component in B at the four epochs plotted against time. For epochs 1--3 
we used the values given in Rioja et al. (\cite{rioja97}). For 1995.9 we 
used AIPS task UVFIT to estimate a separation from the B visibility data 
directly, in order to follow the methodology used for the other epochs as 
closely as possible; the value obtained was 1.895 mas.
The slope from a least-squares fit corresponds to an expansion rate  
of $13.0~\pm~0.7~\mu$as~yr$^{-1}$.
In the standard picture of extragalactic radio sources, the ``core'' is
stationary, so this corresponds to an outward expansion of the
reference component along PA 127\degr.


The rms of the fit (8~$\mu$as) is again surprisingly low, implying
typical errors in the separation measurements at each epoch 
(both within the B structure and between the reference points)
of only about 10--12$~\mu$as along the direction of the B source axis.
This is considerably less than the estimate of position separation 
errors given in Sect. 4.3. 

\subsection{Relative proper motion }

The analysis presented in the previous sections demonstrate
clearly that the chosen reference component within quasar B is
unsuitable for use as a marker for tracing any relative proper motion
between quasars A and B. The value of its expansion velocity derived 
in Sect. 5.4 appears to differ significantly from that deduced by 
vector-decomposition in Sect. 5.3. Although the difference between 
these estimates, if real, could be interpreted as motion of the core of B  
at a rate of $\sim 4~\mu$as~yr$^{-1}$, this is not a conclusive result
since differences of this order arise from choosing different values of 
PA for the motion in A in the vector decomposition method.

\begin{figure}[h]
\centerline{
{\psfig{figure=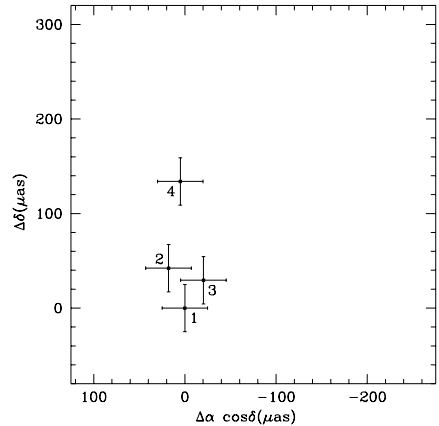,width=6.cm}}
}
\caption{Separation between the cores of A and B (with respect to epoch 1),
derived by correcting the A-B reference point separation measurements with the
core-reference separations measured in the B hybrid maps.
Plotted error bars correspond to $25~\mu$as.
\label{fig7}} 
\end{figure}

A more suitable tracer of relative proper motion between the quasars is 
the variation of the separation between the cores of A and B.
We have used the separations between the core and reference component
measured in the B map at each epoch, and the astrometric
separations between A and B, to calculate the separations between the
A and B cores at each epoch; these are plotted in Fig.~\ref{fig7}.
The area occupied by the points defines an upper limit of 
$\sim 10~\mu$as~yr$^{-1}$ for any relative proper motion between 
the A and B cores, and hence between the quasars themselves, during
the period of nearly 15 years for which the separation has been
monitored with VLBI.
The limit seems to be set by the relatively large deviation of the
1995.9 epoch point in the direction of the A source axis, presumably
arising from the difficulty in defining the reference point at the
A ``core'' from epoch to epoch. 


\subsection{Possible ``core'' motions ? } 

\begin{figure*}
\centerline{
{\psfig{figure=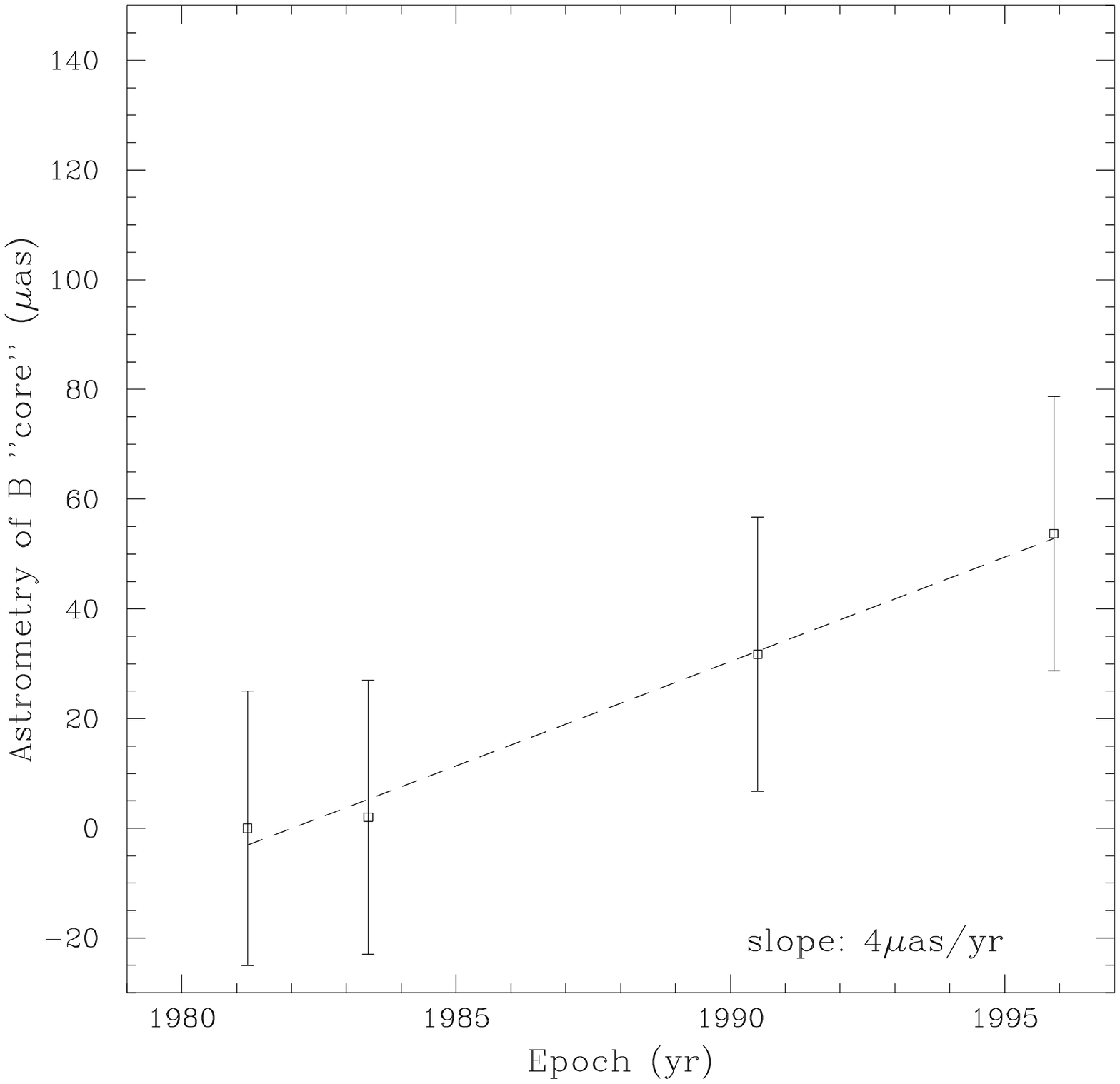,width=6.cm}}
\put (-140,140) {\bf {a)}}
{\psfig{figure=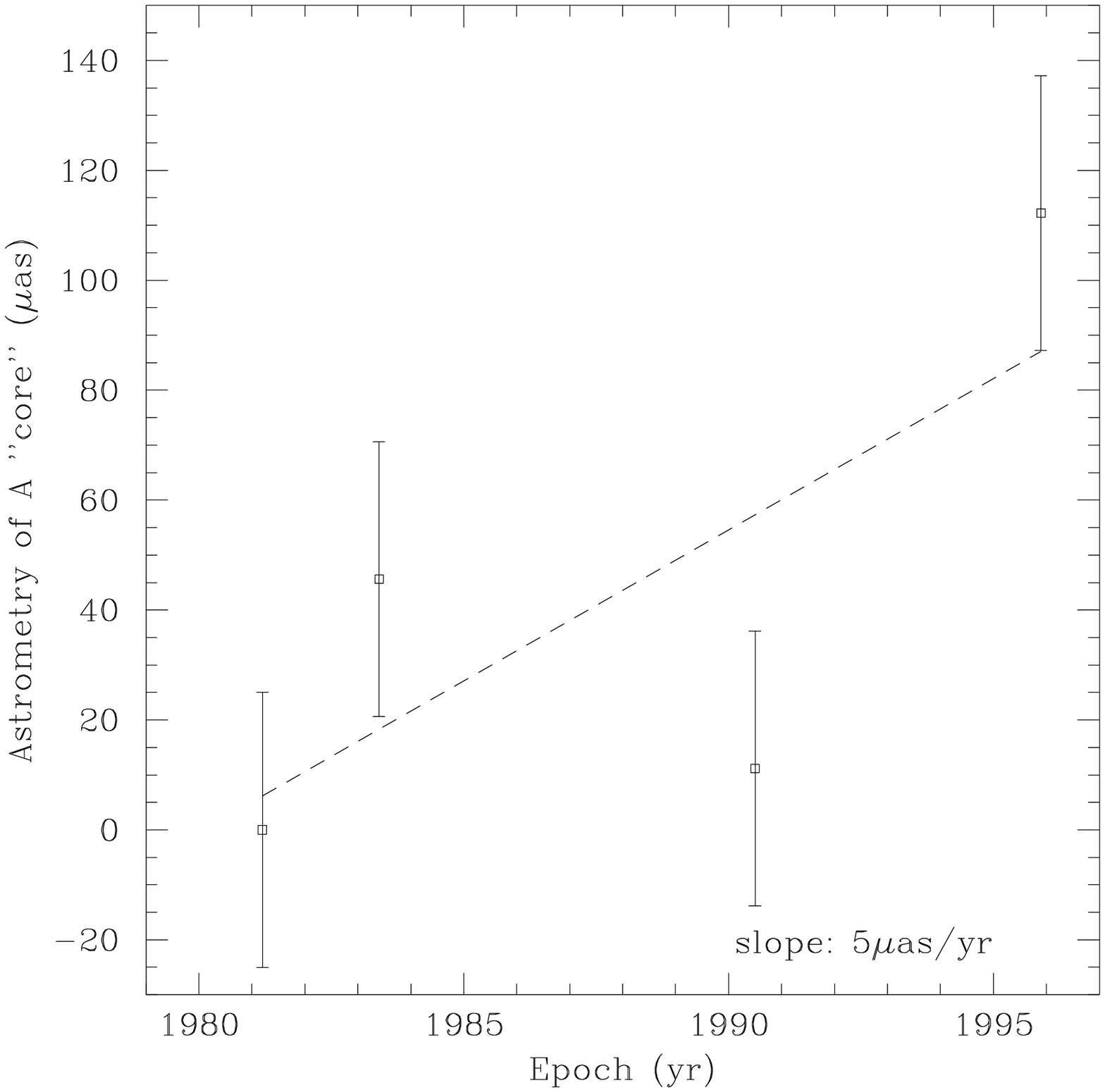,width=6.cm}}
\put (-130,140) {\bf {b)}}
}
\caption{Residual motions of the cores using the deconvolution approach.
 {\bf a} Displacements of B core along PA 127\degr; 
{\bf b} Displacements of A core along PA 25\degr. 
Plotted error bars correspond to $25~\mu$as.
\label{fig8}} 
\end{figure*}

Finally, we investigate any possible residual motions of the ``cores'' 
in A and B. The most likely causes of any such apparent motions are 
changes in the relative brightness or positions of features in the 
source structures at a resolution below that of the maps.
One might expect that these, too, would produce effects predominantly
along the source axis directions.
We therefore used the vector deconvolution method on the plot of
core-core separation with time to study displacements of the cores
along their source axis directions.
Fig.~\ref{fig8}a and b show plots of the separated contributions from 
B and A, for an assumed A source axis PA 25\degr.
The displacements for the B core seem to increase systematically.
The fitted rate is $3.8~\pm~0.3~\mu$as~yr$^{-1}$, indicating a possible 
slow outward motion. The displacements for the A core do not seem to 
vary systematically - the fitted slope is $5.5~\pm~3.6~\mu$as~yr$^{-1}$.
Here the scatter is considerably larger, reflecting both the difficulties of
defining the reference point along the A core-jet axis, and
also, perhaps, real ``jitter'' of the position of the peak due to variations 
in the ``core'' substructure. These plots indicate the level of stability 
of the individual core positions; the fits represent realistic upper limits 
to any possible systematic core motion in the A and B quasars along their 
source axis directions.


\section{Conclusions} 

The series of astrometric VLBI measurements of the separation between 
the quasar pair 1038+528 A and B, spanning
nearly 15 years, provides excellent material for investigating 
the relative proper motion of two extragalactic radio sources and
the positional stability of their cores.
The changes measured in the separations between quasars A and B 
at 3.6~cm are dominated by the motion of the reference feature in quasar B. 
These astrometric results, and measurements in the hybrid maps of B are
compatible with an expansion rate for the B reference component of 
13--17$~\mu$as~yr$^{-1}$.
At a redshift of 2.296 this translates to an apparent
transverse velocity of 0.55--0.70~c~h$^{-1}$.
We note that this is an order of magnitude smaller than the more
typical superluminal velocities seen in many quasars; it is a 
rare example of a subluminal velocity measured for a knot in a quasar jet.

After correcting for the motion of the reference component in B,
we can put a conservative upper bound to any relative proper
motion between the quasars of $10~\mu$as~yr$^{-1}$.
Despite the increase in temporal baseline, this upper bound is no
better than that given by Rioja et al. (\cite{rioja97}).
Its value is related to the difficulty in reproducing a stable reference
position along the A source axis near its ``core''.

Theories in which the redshifts of quasars do not indicate cosmological
distances, and in which quasars are ``local'' and have high Doppler
redshifts (e.g. Narlikar \& Subramanian \cite{narlikar83}) are incompatible 
with our measured upper-limit to relative proper motion. Quasars at 
100 Mpc distance moving at relativistic speeds would have proper motions 
of the order of $600~\mu$as~yr$^{-1}$, nearly 2 orders of magnitude 
greater than our limit. Assuming cosmological distances,
our limit corresponds to apparent transverse velocities of 
0.43~c~h$^{-1}$ and 0.22~c~h$^{-1}$ at the redshifts of the B and A
quasars, respectively.

We have investigated the way in which the definition of   
reference points in a map may be only loosely ``fixed'' to the radio
source structure, especially when the latter is strongly asymmetric.
We have also developed an alternative analysis route - Hybrid Double
Mapping - for imaging both sources of a close pair simultaneously, and
at the same time preserving their relative astrometric information
in a single map.

The surprisingly low rms from the fits of linear expansion in quasar B,
and the discrepancy between the two estimates   
($16.9~\pm~0.6~\mu$as~yr$^{-1}$ from the astrometric measurements and
 $ 13.0~\pm~0.7~\mu$as~yr$^{-1}$ from the hybrid maps)
are suggestive of (but do not prove) a residual motion of the core 
in quasar B; our decomposition along the source axis direction gives a fit
of $3.8~\pm~0.3~\mu$as~yr$^{-1}$, corresponding to an apparent transverse 
velocity of 0.17~c~h$^{-1}$. If real, this might indicate a steady change 
in physical conditions at the base of the jet, or perhaps the emergence 
of a new knot component moving outwards with a velocity similar to the 
reference component, but as yet unresolved by our 0.5 mas beam.
In this regard, it is interesting to note the slight extension of
the core of quasar B in PA 132\degr given by the Gaussian model fit.

The low rms derived from fits to the expansion of the B reference component 
indicate that we have been overly conservative in our estimate of 
$18~\mu$as for the error in reference point positions. Errors at least 
2 times smaller are implied, corresponding to a sixtieth of the beamwidth. 
It is interesting to note that such small errors are also implied in the 
work of Owsianik \& Conway (\cite{owsianik98}), where the low scatter in 
the plot of expansion of the CSO source \object{0710+439} allows an 
expansion rate of $14.1~\pm~1.6~\mu$as~yr$^{-1}$ to be determined.

There are no obvious systematic motions within quasar A, but the ``noise'' 
in the estimates of position along its axis are much larger.
This noise, along with any associated underlying changes in source
substructure, provides a fundamental limit to estimates of
any systematic core motion in A. Improvements on the estimates of (or upper
bounds to) the relative motion between the quasars, or of the
individual motion of the A core, will require a considerable
increase in the temporal baseline of VLBI monitoring.

\newpage

\appendix 

\section {Hybrid Double Mappping (HDM)}

{\bf A.1 Principle of Hybrid Double Mapping} \\

\noindent
The visibility function, $V$, measured at time $t$, on a baseline between
antennas $i$ and $j$, is represented by a complex function with amplitude $A$, 
phase $P$:

\[ V(i,j) = A * e^{i[\phi]}(t) \]

\noindent
For this analysis it is convenient to indentify 3 contributions to
the visibility phase:

\[ \phi(i,j) = \phi_{s}(u,v) + \phi_{p}(u,v) + \phi_{m}(t) \]

\noindent
where  

\begin{description} 
\item [$\phi_s$] is due to source structure, evaluated w.r.t. a reference
position for the source. \\

  \item [$\phi_p$] is due to any offset of the 
true source position from the reference position. \\

Both $\phi_s$ and 
$\phi_p$ are functions of the resolution coordinates, $u$ and $v$, at time 
$t$. \\

\item [$\phi_m$] is due to inaccuracies in the correlator model calculation of
         the interferometer geometry and the signal propagation delays
         in the ionosphere, troposphere and receiving system; it is an  
         unknown function of time. \\
         This term can be represented by the difference of two
         ''antenna-based'' phases, $\theta_i$ and $\theta_j$, since it can 
         be related to the 
         difference in signal arrival times at the two sites.
         (This analysis is a simplification which ignores possible
         ''non-closing'' instrumental baseline phase terms arising from
         e.g. un-matched bandpasses and polarisation impurities.) 
\end{description}

\[ \phi(i,j) = \phi_s(u,v) + \phi_p(u,v) + \theta_i(t) - \theta_j(t) \]

\noindent
In conventional hybrid mapping, an iterative procedure is used
to separate out the antenna-based phase terms from the ''source'' terms;
the latter must produce a consistent and physically plausible source
structure after Fourier transformation of the corrected visibility:

\[ \underbrace {A * e^{i[\phi_s+\phi_p]}(u,v)}_{corrected \: visibility} =  
V(i,j) * \underbrace {e^{-i[\theta_i(t)-\theta_j(t)]}}_{antenna \: phase \: terms} \] 

\noindent
However, the position offset term, $\phi_p$, can also be expressed as a difference 
in wavefront arrival times at the 2 antennas and so it is also
''absorbed'' in antenna phase terms $\theta_i'$, $\theta_j'$; 
the ''absolute'' position information is lost:

\[ \underbrace {A * e^{i[\phi_s]}(u,v)}_{corrected \: visibility} =   
V(i,j)  *  \underbrace {e^{-i[\theta_i'(t)-\theta_j'(t)]}}_{antenna \: phase
\: terms} \]

\noindent
In Hybrid Double Mapping (HDM), the visibility functions of two 
sources observed simultaneously are added. For a close source pair, we 
make the same assumption as for conventional phase-referencing - that the 
model error phase terms are essentially the same for both sources. We make 
a further assumption that the $u,v$ coordinates are also essentially the same
for both sources, for each baseline and time. The visibility sum, $V^1 + V^2$,
can then be re-written: \\

\[ \underbrace {(A^1 * e^{i[\phi^1_{s}]} + A^2 * e^{i[\phi^2_{s}+(\phi^2_{p}
-\phi^1_{p})]})(u,v)}_{corrected \: visibility \: sum} = \]
\[ \;\;\; \vspace*{2cm} = V^{sum}(i,j) * \underbrace {e^{-i[\theta_i'(t)-\theta_j'(t)]}}_{antenna \:
 phase \: terms} \]


\noindent
This may be recognised as the visibility function of a ''composite''
source consisting of the sum of the brightness distributions of sources
1 and 2, with antenna-based phase error terms $\theta_i'$, $\theta_j'$, 
as before.
The HDM method consists of performing the normal hybrid mapping
procedure with the visibility sum, resulting in the separation of the
antenna-based errors, and a physically plausible map of the sum of
the two source brightness distrubutions. An important point is that,
whereas the {\it origin} of the map of the composite source is arbitrary (as
it depends on the position of the starting model), the {\it separation} of
the two source brightness distributions within the composite map
(determined by $\phi^2_{p}-\phi^1_{p}$) is fixed during the phase 
separation procedure,
and is equal to the {\it difference} of the errors in the two source
positions used for correlation. We call this the ''residual separation''. \\

\noindent
{\bf A.2 Practical aspects} \\

\noindent
There are some practical aspects to be considered. If the source
coordinates used in the correlator model are very precise, then the
residual separation may be less than the interferometer beamwidth, and
the two source distributions will lie on top of each other. 
In this case it is desireable to
introduce an artificial position offset into one of the
source visibility functions before forming the visibility sum, to
ensure that the two source reference features are well separated in the
HDM map. One should also arrange that the peak of one source
does not lie on the sidelobe response of the other in the "dirty" map,
as this may degrade the CLEAN deconvolution process in the mapping step.\\

\noindent
      Another important consideration is that the time-averaged samples
of the summed visibility function contain equal contributions from both
source visibility functions. When both sources are observed
simultaneously this will normally be the case, except when different
amounts of data are lost in the two separate correlator passes needed
for the two source positions.  It is important to edit the data sets
carefully to fulfil this condition. \\

\noindent
      The range of validity of the assumption that the $(u,v)$
coordinates for the two sources are the same depends on the ''dilution
factor'', i.e. the reciprocal of the source separation, measured in
radians. The $(u,v)$ value assigned to the summed visibility will be
incorrect for either source by roughly 1 part in the dilution factor
(roughly 1 in 6000 for 1038+528A,B).  
This is equivalent to having source visibility phase errors of this
order, and thus limits the size of an HDM map to be less than the
beamwidth times the dilution factor; the residual separation should be
much smaller than this value.\\

\noindent
In the actual analysis used in this work, we first made a rough 
correction to the phase of the summed visibility of 1038+528 A + B, using
the antenna phase and phase derivative errors from fringe-fitting
1038+528A using a point source model. However, there is no reason why
one should not fringe-fit the summed visibility function directly. \\

\noindent
{\bf A.3 Applications} \\

\noindent
The HDM method can in principle be applied
whenever two (or more !) radio sources are observed simultaneously,
but are correlated at separate field centres; however, 
they must be close enough so that the conditions of same $(u,v)$
coverage and same correlator model errors apply.
The method uses the structures of BOTH sources simultaneously to
separate out the antenna phase errors, as opposed to a single source in  
simple hybrid mapping.
If both sources are strong (as with 1038+528 A and B),
constraining the (single) antenna phase solutions with two structures
should lead to a more rigorous and robust separation between the
source and antenna phase terms.
One field of application is in high resolution VLBI imaging of
gravitational lens systems with wide image separations (e.g.  
images A and B of QSO 0957+561 with 6.1 arcsec separation)
where preserving the necessary wide field-of-view from a single 
correlation may result in inconveniently large data sets.
When one source is very weak, however,
there is probably little to be gained over normal hybrid mapping. \\

\noindent
For relative astrometry studies (as described in this paper),
the HDM method has some advantages over conventional phase-reference 
mapping and explicit phase-differencing methods.
In phase-differencing astrometry, separate hybrid maps must be made of
both sources to correct for source structural phase terms and 
the antenna phase errors are NOT constrained to be the same. Imperfect 
separation between source and antenna phase terms can increase the noise 
on the differenced phase, as well as lead to possible systematic errors.
In phase-reference astrometry, only one of the sources is used to solve
for the antenna phase terms; imperfect separation can lead to extra
phase noise in the phase-referenced visibility of the ''target'' source.
In HDM we use both source
structures simultaneously to separate the (common) antenna error
phases from that of a single "structure" in which the 
reference points of the two sources are spaced by the residual separation.\\

\noindent
When the separation between the two sources of a pair exceeds
the telescope primary beamwidths, astrometric and phase-reference 
observations  
must involve switching between the sources, and the visibility phase of at  
least one of the sources must normally be interpolated in the observing gap. 
The condition that must be fulfilled for HDM to work in
this case is that an equal number of observations of both sources must
be added to form an average visibility function for the length of the
''solution interval'' in the phase self-calibration step of HDM.
This length is generally limited by the coherence time of the
atmosphere, and would imply a very fast switching cycle in most cases.
Another application for HDM could be in the the analysis of
''cluster-cluster'' VLBI (see e.g. Rioja et al. \cite{rioja97b}), 
in which two or more sources are observed simultaneously on VLBI baselines 
by using more than one telescope at each site.

\begin{acknowledgements}

We thank Dave Graham for help with the observations in Effelsberg,
Tony Beasley for asistance at the VLBA, and Mark Reid for useful comments
on the text. M.J.R. wishes to acknowledge support for this research by the
European Union under contract CHGECT920011.
The National Radio Astronomy Observatory is a facility
of the National Science Foundation operated under cooperative
agreement by Associated Universities, Inc.

\end{acknowledgements}

{}

\end{document}